\documentclass[10pt]{article}
\usepackage{epsf} 
\usepackage{amsmath}
\usepackage{amssymb}
\usepackage{epsfig}
\usepackage{latexsym}
\usepackage{amsfonts}
\usepackage{graphicx}
\usepackage{cite}
\usepackage{color}
\begin{document}
\parindent 0mm 
\setlength{\parskip}{\baselineskip} 
\thispagestyle{empty}
\pagenumbering{arabic} 
\setcounter{page}{1}
%\mbox{ }
\begin{center}
{\Large {\bf THEORETICAL DETERMINATION OF THE HADRONIC\\
\vspace{4mm}
 g-2 OF THE MUON}}
\\
\end{center}
%\vspace{.05cm}
%{\footnote{{\LARGE {\footnotesize Speaker.}}}}
\begin{center}
C. A. DOMINGUEZ $^{(a),(b)}$ $^{*}${\footnote{{\LARGE {\footnotesize Speaker.}}}},  K. SCHILCHER $^{(a),(b),(c)}$ $^{\dagger}$ and H. SPIESBERGER $^{(a),(c)}$ $^{\dagger \dagger}$\\ 
\end{center}
\begin{center}
$^{*}$ cesareo.dominguez@uct.ac.za 
\end{center}
\begin{center}
$^{\dagger}$ karl.schilcher@uni-mainz.de
\end{center}
\begin{center}
$^{\dagger\dagger}$ spiesber@uni-mainz.de
\end{center}
\begin{center}
{\it $^{(a)}$ Centre for Theoretical and Mathematical Physics, 
and Department of Physics, University of
Cape Town, Rondebosch 7700, South Africa}
\end{center}
\begin{center}
{\it $^{(b)}$ National Institute of Theoretical Physics, Private Bag XI, Matieland 7602,  South Africa}
\end{center}
\begin{center}
{\it $^{(c)}$ PRISMA Cluster of Excellence, Institut f\"{u}r Physik, 
Johannes Gutenberg-Universit\"{a}t, D-55099 Mainz, Germany}
\end{center}
\begin {abstract}
\noindent
An approach is discussed on the determination of the leading order hadronic contribution to the muon anomaly, $a_\mu^{HAD}$, based entirely on theory. This method makes no use of $e^+ e^-$ annihilation data, a likely source of the current discrepancy between theory and experiment beyond the $3\, \sigma$ level. What this method requires is essentially knowledge of the first derivative of the vector current correlator at zero-momentum. In the heavy-quark sector this is obtained from the well known heavy quark expansion in perturbative QCD, leading to values of $a_\mu^{HAD}$ in the charm- and bottom-quark region which were fully confirmed by later lattice QCD (LQCD) results. In the light-quark sector, using recent preliminary LQCD results for the first derivative of the vector current correlator at zero-momentum leads to the value $a_\mu^{HAD} = (729 - 871)\, \times\,10^{-10}$, which is significantly larger than values obtained from using $e^+ e^-$ data.
A separate approach based on the operator product expansion (OPE), and designed to quench the contribution of these data, reduces the discrepancy by at least 40\%. In addition, it exposes a tension between the OPE and  $e^+ e^-$  data, thus suggesting the blame for the discrepancy on the latter.
\end{abstract}

\section*{1. Introduction}
In this talk I first discuss  a novel method \cite{amu1} to determine the leading order hadronic contribution to $(g-2)$ of the muon,  $a_\mu^{HAD}$ \cite{review}. This allows for a purely theoretical determination based on QCD, thus avoiding the use of $e^+ e^-$ annihilation data, which are affected by many uncertainties. The method consists in replacing the well known integration kernel $K(s)$, entering the expression for the anomaly $a_\mu^{HAD}$, by a fit function having simple poles in the complex square energy $s-$ plane at $s=0$. Invoking Cauchy's residue theorem in this plane, $a_\mu^{HAD}$ is fixed by perturbative QCD (PQCD) and the residues of the poles. In the heavy-quark sector (charm and bottom) the well known PQCD expansion of the vector current correlator around $s=0$ fixes these residues. The predictions from this method in the charm- and  bottom-quark regions were later fully confirmed by Lattice QCD (LQCD) determinations\cite{LQCD1}-\cite{LQCD3}. Regarding the leading contribution from the light-quark sector,  the derivatives of the vector current correlator at $s=0$ were estimated in the framework of a Large $N_c$-QCD model of the pion form factor \cite{amu1}. While the result was fully consistent with the expectation from the Standard Model (SM), it clearly remains a model-dependent result.  Current preliminary results from LQCD for the first derivative of the vector current correlator at the origin \cite{Wittig}, which dominates the result by roughly an order of magnitude, lead to a complete theoretical prediction of the anomaly. The result is substantially larger than values obtained using $e^+e^-$ data. If the LQCD results from \cite{Wittig} were to be confirmed, then it would be possible to understand the muon $(g-2)$ value within the Standard Model.\\ 

Further support for the view that the culprit in the $g-2$ saga could  be the $e^+ e^-$ data (in the light-quark sector) is provided by a second approach to the determination of the anomaly \cite{amu2}, also to be discussed here. This consists in using the operator product expansion (OPE) of current correlators at short distances, plus a finite energy QCD sum rule (FESR) designed to quench the role of the $e^+ e^-$  data. A clear tension between the OPE and  $e^+ e^-$ data was identified, suggesting a cross-section deficit in these data. While not solving completely the discrepancy between theory and experiment, this approach reduces it by at least 40\%.\\
Since QCD FESR and the OPE, which rely on quark-hadron duality, have lately been questioned  \cite{DV}, a specific model of duality violations requiring six free-parameters \cite{Cata}  was used in \cite{amu2} in order to check this issue. Our results \cite{amu2} were fully consistent with no duality violations taking place, at least in this application.

\section*{2. Theoretical Determination of the Leading Order  $a_\mu^{HAD}$: The Method}

The standard expression of the (lowest order) muon anomaly is given by
%Eq.1
\begin{equation}
a_{\mu}^{HAD}=\;\frac{\alpha_{EM}^{2}}{3\,\pi^{2}}\,\int_{s_{th}=m_{\pi}^2}^{\infty
}\,\frac{ds}{s}\;K(s)\;R(s)\;,\label{AMUH1}%
\end{equation}
where $\alpha_{EM}$ is the electromagnetic coupling, and the standard $R$
ratio is $R(s)=3\,\sum_{f}\,Q_{f}^{2}\left[  8\,\pi\,\mbox{Im}\,\Pi
(s)\right]  $.  $\Pi(s)$ is the vector
current correlator 
%Eq.2
\begin{eqnarray}
\Pi_{\mu\nu} (q^2) &=& i \int d^4x\,  e^{iqx} \langle 0| T \left(j^{\text{\,EM}}_{\mu}(x)\, j^{\text{\,EM}}_{\nu}(0)\right)|0\rangle \nonumber\\ 
&=& (q_\mu q_\nu - q^2 g_{\mu\nu}) \Pi(q^2)\;, \label{eq:correlator}
\end{eqnarray}
where the electromagnetic current is  $j^{\text{EM}}_\mu(x)=\sum_f Q_f \bar{q}_f(x) \gamma_\mu q_f(x)$, with the sum running over   quark flavours, and $Q_f$ are the quark charges. The function $\Pi(s)$ is normalized to $6\,\pi\,\mbox{Im}\,\Pi(s)=1+\alpha_{s}%
/\pi+\cdots$.
The integration kernel $K(s)$, at leading order, in Eq.(\ref{AMUH1}) is given by \cite{deRK}
%Eq.3
\begin{equation}
K(s)=\int_{0}^{1}\,dx\,\frac{x^{2}(1-x)}{x^{2}+\frac{s}{m_{\mu}^{2}}
(1-x)}\;, \label{K}
\end{equation}
where $m_{\mu}$ is the muon mass. For convenience  one splits $a_\mu^{HAD}$  into the contributions from the three quark-mass regions labelled by the quark flavours $(u,d,s)$, $c$, and $b$, i.e.
%Eq.4
\begin{equation}
a_{\mu}^{HAD}=a_{\mu}^{HAD}|_{uds}+a_{\mu}^{HAD}|_{c}+a_{\mu}^{HAD}
|_{b} \;. 
\end{equation}
In order to be able to determine each one of these contributions entirely from theory, i.e. avoiding the use of $e^+ e^-$ data, I first replace the kernel $K(s)$, Eq.(\ref{K}), by  fit functions $K_i(s)$, with $i=1,2,3$ corresponding to the three quark flavour regions $(u,d,s)$, $c$, and $b$, respectively. The  $K_i(s)$ are chosen as meromorphic functions with simple poles at the origin. This allows for the use of Cauchy's theorem in the complex square energy $s$-plane, relating information on the positive real $s$-axis to that around a circle of radius $s=|s_0|$, large enough for PQCD to be valid. The determination is completed by adding the line integral in the region $s\in \, s_0 - \infty$, for which the exact integration kernel $K(s)$, Eq.(3), can now be used.\\
Starting with the light-quark sector, the appropriate fit function $K_1(s)$ in the interval $s_{th} \leq s \leq s_0$ is of the form
%Eq.5
\begin{equation}
K_{1}(s)=a_{0}\,s+\,\sum_{n=1}\frac{a_{n}}{s^{n}}\;,\label{K1}
\end{equation}
with coefficients determined by minimizing the chi-squared. The upper limit
$s_{0}$ is below the charm threshold. Invoking Cauchy's theorem one obtains
%Eq.6
\begin{equation}
\int_{s_{th}}^{s_0} \frac{ds}{s}  K_{1}(s) \, \frac{1}{\pi} \, \mbox{Im} \,\Pi_{uds}(s)  = 
 \mbox{Res} \left[ \Pi_{uds}(s) \frac{K_{1}(s)}{s}\,\right]_{s=0}  
-  \frac{1}{2 \pi i} \oint_{|s|=s_0}\frac{ds}{s} \; K_{1}(s)\, \Pi_{uds}(s)  \;, \label{CAU}
\end{equation}
where the integral on the right hand side, around the circle of radius
$s_{0}\simeq(1.8\;\mbox{GeV})^{2}$, is computed using PQCD in the light-quark sector. This is known up to five-loop level \cite{5L}. The contour integration can be performed using fixed order perturbation theory (FOPT) or, alternatively, contour improved perturbation theory (CIPT). In this application both give essentially the same answer. The final expression for the anomaly becomes
%Eq.7
\begin{eqnarray}
a_\mu^{HAD}|_{uds} &=& 8 \alpha_{EM}^2 \sum_{i=u,d,s} Q_i^2 \left\{ \mbox{Res} \left[ \Pi_{uds}(s) \frac{K_{1}(s)}{s}\right]_{s=0}  
-  \frac{1}{2 \pi i} \oint_{|s|=s_0}\frac{ds}{s} \; K_{1}(s) \; \Pi_{uds}(s)|_{PQCD}  
\nonumber \right. \\ 
&+& \left.
\int_{s_0}^{\infty} \, \frac{ds}{s} \; K(s) \; \frac{1}{\pi} \, \mbox{Im}\, \Pi_{uds}(s)|_{PQCD}\right\}, \label{AMUL}
\end{eqnarray}
where the last integral above involves the exact integration kernel $K(s)$ and PQCD is used for the spectral function. The threshold for PQCD is chosen as $s_0 \simeq \, {(1.8 \;\mbox{GeV})^2}$. This allows for a fair comparison with determinations based entirely on $e^+ e^-$ data, and  it is  supported by BES data \cite{BES} suggesting $s_0 \simeq \, {(2.0 \;\mbox{GeV})^2}$.\\
The contribution from the heavy-quark sector (charm and bottom) is obtained from Eq.(\ref{AMUL}) after obvious replacements, and substituting the fit kernel $K_1(s)$ by fit kernels $K_2(s))$ and $K_3(s)$ for the charm- and bottom-quark regions, respectively.
%Fig.1
\begin{figure}
	[ht]
	\begin{center}
		\includegraphics[height=1.5in]{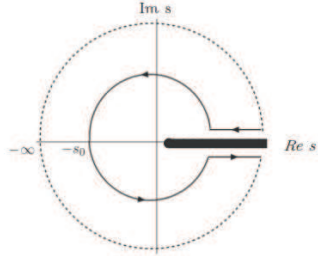}
		\caption{\footnotesize{The squared energy $s$-plane used in Cauchy's theorem.}}
	\end{center}
\end{figure}
%Fig 2 
\begin{figure}[ptb]
\begin{center}
\includegraphics[height=2.5in]{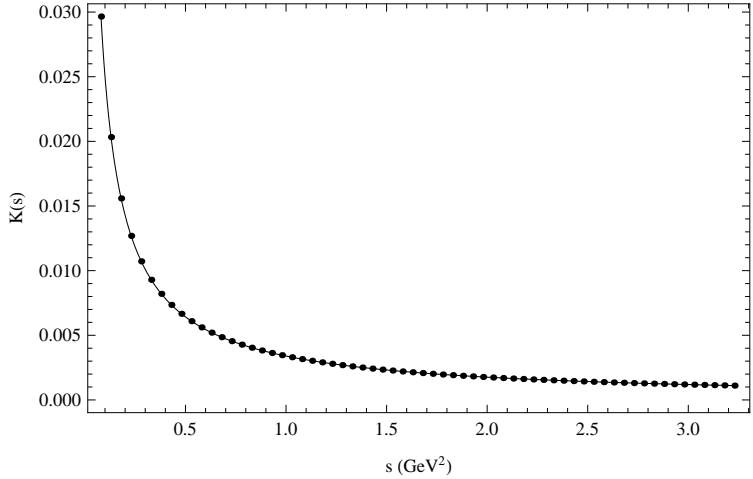}\caption{{\footnotesize
{The exact kernel $K(s)$, Eq.(\ref{K}) (solid line) together with the fit in the
light-quark region, Eq.(\ref{K1N}), (solid circles). }}}%
\end{center}
\label{figure2}%
\end{figure}
The optimal fit function, Eq.\ref{K1}, resulting in the lowest chi-squared was found to be
%Eq.8
\begin{equation}
K_{1}(s) = 2.257\times 10^{-5} s + 3.482\times 10^{-3} s^{-1}
- 1.467\times 10^{-4} s^{-2} 
+ 4.722\times 10^{-6} s^{-3} \;, \label{K1N}
\end{equation}
where $s$ is expressed in $\mbox{GeV}^{2}$, and the numerical coefficients
have the appropriate units to render $K_{1}(s)$ dimensionless. Figure 1 shows
the exact kernel $K(s)$ in Eq.(\ref{AMUH1}) (solid curve) together with the fit $K_{1}(s)$ as in Eq.(\ref{K1N}) (solid dots). The relative difference between the two curves lies in the range $0-1\%$ in the low energy region, where it contributes the most. In other words, there is essentially no difference in the result for $a_{\mu}^{HAD}|_{uds}$ in this energy region if one uses the exact kernel, Eq.(\ref{K}), or the fit kernel, Eq.(\ref{K1}).
This functional form for the kernel will require knowledge of the first three derivatives of the vector correlator at the origin. However, as discussed later, the first derivative dominates over the second and the third by one and by two orders of magnitude, respectively. This is due to the relative size of the fit coefficients in Eq.(\ref {K1N}).\\
Proceeding to the heavy-quark sector, the fit to the kernel $K(s)$, Eq.(\ref{K}), named $K_2(s)$ in the charm-quark region, $s_1 \simeq M_{J/\psi}^{2} \leq s \leq s_2 \simeq(5.0\,\mbox{GeV})^{2}$, is
given by
%Eq.9
\begin{equation}
K_{2}(s)=\frac{a_1}{s}\;+\;\frac{a_2}{s^2}\;,
\end{equation}
where $a_{1}=0.003712\;\mbox{GeV}^{2}$ and $a_{2}=-0.0005122\;\mbox{GeV}^{4}$. This function
provides an excellent fit, as it  differs from the exact kernel
$K(s)$ by less than $0.02\%$. The expression for the anomaly is now given by Eq.(\ref{AMUL}), with the obvious replacements. In the bottom-quark region, the corresponding fit function is now given by 
%Eq.10
\begin{equation}
K_{3}
(s)=0.003719\,\mbox{GeV}^{2}\,s^{-1}-0.0007637\,\mbox{GeV}^{4}\,s^{-2}.
\end{equation}
This  kernel differs from the exact kernel, $K(s)$ by less than 0.0005 \% in the range $M_{\Upsilon}^{2}\leq s\leq(12\,\mbox{GeV})^{2}$.\\
An important difference between the light-quark and the heavy-quark sector, is that in the latter the vector correlator and  its derivatives at $s=0$  can be computed in QCD, to wit. The Taylor series heavy-quark expansion of the correlator around the origin is
%Eq.11
\begin{equation}
\Pi_{c,b}(s)|_{PQCD}=\frac{3}
{32\pi^{2}}\,Q_{c,b}^{2}\,\sum_{n\geq0}\bar{C}_{n}z^{n}\;,
\end{equation}
where $z=s/(4\overline{m}_{c,b}^{2})$. Here $\overline{m}_{c,b}\equiv\overline{m}_{c,b}(\mu)$ is the charm (bottom)-quark mass in the $\overline{\text{MS}}$-scheme at a renormalization scale $\mu$. The coefficients $\bar{C}_{n}$ up to $n=30$ are known at three-loop level \cite{boughezal2006a}-\cite{maier2008a}. At four-loop level $\bar{C}_{0}$ and $\bar{C}_{1}$ were determined in \cite{boughezal2006a}-\cite{chetyrkin2006},
$\bar{C}_{2}$ in \cite{maier2008a} and $\bar{C}_{3}$ in \cite{maier2010}.  Due
to the s-dependence of $K_{2,3}(s)$ no coefficients $\bar{C} 
_{4}$ and higher contribute to $\text{Res}[\Pi_{c,b}(s)\,p(s),s=0]$.
\section*{3 Results}
To begin, I concentrate on the two integrals in Eq.(\ref{AMUL}), and consider each of the three quark-sections separately, i.e. ($u,d,s$), $c$, and $b$. In the first section $s_{th} = m_\pi^2$,  the radius of the Cauchy circle is $s_0 \simeq (1.8 \; \mbox{GeV})^2$, and the lower limit of the line integral is also this radius. In the second section the threshold is $s_{th}\equiv s_1 \simeq M_{J/\psi}^2$, and the radius/lower limit of the line integral is $s_0 \equiv s_2 \simeq (5.0\; \mbox{GeV})^2$. For the third section $s_{th}\equiv s_3 \simeq M_{\Upsilon}^2$, and
 $s_0 \equiv s_4 \simeq (12.0\; \mbox{GeV})^2$. The PQCD expansion of the vector correlator needed in the contour integral is of the form
 %Eq.12
 \begin{equation}
 \Pi_{\text{PQCD}}(s)=\sum_{n=0}^{\infty}\left(  \frac{\alpha_{s}(\mu^{2}%
 )}{\pi}\right)  ^{n}\Pi^{(n)}(s)\;,
 \end{equation}
where 
 %Eq.13
 \begin{equation}
 \Pi^{(n)}(s)=\sum_{i=0}^{\infty
 }\left(  \frac{\overline{m}^{2}}{s}\right)  ^{i}\Pi_{i}^{(n)}\;.
 \end{equation}
  The complete analytical result in PQCD up to $\mathcal{O}(\alpha_{s}^{2},(\overline{m}
 ^{2}/s)^{6})$ is given in \cite{Chetyrkin1997}, with new results up to order ${\cal{O}}(\alpha_s^2 (\overline{m}^2/s)^{30})$ obtained recently \cite{MAIER2}.
 There are also exact results
 for $\Pi_{0}^{(3)}$ and $\Pi_{1}^{(3)}$ from \cite{Baikov2009}, while $\Pi
 _{2}^{(3)}$ is known up to a constant term \cite{Chetyrkin2000b}. This
 constant term does not contribute to the contour integral due to the
 $s$-dependence of $K_{2}(s)$. Finally, at five-loop level the
 full logarithmic terms in $\Pi_{0}^{(4)}$ and $\Pi_{1}^{(4)}$ are known from
 \cite{Baikov2008} and \cite{Baikov2004}, respectively. With this information, the three contour integrals in FOPT are
 %Eq.14
 \begin{eqnarray}
 \frac{1}{2 \pi i} \oint\frac{ds}{s} \; K_{n}(s) \; \Pi_{q}(s)|_{PQCD} \;  = \;\left\{ 
 \begin{array}{lcl}
 135.3 (6) \times 10^{-7} \\
 \,\, \,20.3 (1) \times 10^{-7} \\
 \, \,\,\,\,\,3.6 (2) \times 10^{-7}  \;,\label{OINT}
 \end{array}\right.
 \end{eqnarray}
for $n=1,2,3$ and $q=(uds),c,b$, respectively. For $n=1$ the result in CIPT is $135.6(6) \, \times\, 10 ^{-7}$, i.e. a 0.2\% difference with FOPT. The results for the line integral in Eq.(\ref{AMUL}), and their equivalent for the charm- and
bottom-quark sectors are
%Eq.15
\begin{eqnarray}
\int_{s_j}^{\infty}  \frac{ds}{s} K(s) \, \frac{1}{\pi} \, \mbox{Im}\, \Pi_{q}(s)|_{PQCD}=\left\{ 
\begin{array}{lcl}
151.8 (1)\times 10^{-7} \\
\,\,\,20.0 (4)\times 10^{-7} \\  
\,\,\,\,\,\,3.4 (2)\times 10^{-7} \label{INT}
\end{array}\right.
\end{eqnarray}
with $j=0,2,4$ for $q=(uds),c,b$, respectively.
Finally, the residues in Eq.(\ref{AMUL}) for the charm- and bottom-quark sector are
%Eq.16
\begin{equation}
\text{Res}\left[  \Pi_{c}(s)|_{PQCD}\frac{K_{2}(s)}{s}\right]_{s=0}
=76.1(5)\,\times10^{-7}\;,\label{RESC}%
\end{equation}
%Eq.17
\begin{equation}
\text{Res}\left[  \Pi_{b}(s)|_{PQCD}\frac{K_{3}(s)}{s}\right]_{s=0}
=6.3\,\times10^{-7}\;,\label{RESB}%
\end{equation}
where the error in Eq.(\ref{RESC}) is due to the uncertainty in $\alpha_{s}$ and to the
truncation of PQCD, while the error in the bottom-quark sector is negligible.
The individual contributions to the anomaly from the charm- and bottom-quark sectors  $a_\mu^{HAD}|_{c,b}$  are
%Eq.18
\begin{equation}
a_{\mu}^{HAD}|_{c}=14.4(1)\times10^{-10}\;, \label{amuc}
\end{equation} 
%Eq.19 
\begin{equation}
a_{\mu}^{HAD}|_{b}=0.29(1)\times10^{-10}\;. \label{amub}
\end{equation}
It is very important to mention that these  results were later fully confirmed by several LQCD determinations \cite{LQCD1}-\cite{LQCD2}. In fact, in the charm-quark sector LQCD finds $a_{\mu}^{HAD}|_{c}=14.42(39) \times 10^{-10}$ from \cite{LQCD1}, and in the bottom-quark sector $a_{\mu}^{HAD}|_{b}=0.271(37)\times10^{-10}$ from \cite{LQCD2}.\\
The  result for the muon anomaly is now
%Eq.20
\begin{equation}
a_{\mu}^{HAD} = \frac{16}{3}\alpha_{EM}^{2}\mbox{Res}\left[  \Pi_{uds}
(s)\frac{K_{1}(s)}{s}\right]  _{s=0} 
+ \, 19.4(2)\times10^{-10}\;,\label{AMUHSF}%
\end{equation}
where the light-quark contribution requires, in principle, the first three derivatives of the vector current correlator at the origin. Near the origin, this correlator is essentially dominated by the pion form factor. Various models of the latter indicate that the first three derivatives are roughly of the same order of magnitude. Given the relative size of the fit parameters in Eq.(\ref {K1N}), this means that thee first derivative is expected to dominate the residue by one to two-orders of magnitude over the second and third derivative, respectively.\\
A preliminary LQCD result \cite{Wittig} for the first derivative of the vector current correlator at the origin is
%Eq.21
\begin{equation}
\Pi^{'}(0)|_{uds} = (0.072 \,- \, 0.087) \; {\mbox{GeV}^{-2}} \;. \label{LQCDuds}
\end{equation}
This range was found by choosing the minimum and maximum values  of the preliminary LQCD results \cite{Wittig} in eleven ensembles for the up-,  down-, and strange-quark contribution.
The residue of the   pole in Eq.(\ref{AMUHSF}) computed using Eq.(\ref{LQCDuds}) becomes
%Eq.22
\begin{equation}
\mbox{Res}\left[  \Pi_{uds}(s)\frac{K_{1}(s)}{s}\right]  _{s=0}^{LQCD}
= (0.25 \,- \, 0.30) \times 10^{-3} \,. \label{lightres}
\end{equation}
This result is compatible with that from the ${\mbox{Dual-QCD}_{\infty}}$ model \cite{amu1}. Substituting Eq.(\ref{lightres}) into Eq.(\ref{AMUHSF}), the contribution to the anomaly from the light-quark sector becomes
%Eq.23
\begin{equation}
a_{\mu}^{HAD}|_{uds} = (710 \, - \, 852) \, \times \, 10^{-10}, \label{uds}
\end{equation}
which is larger than the LQCD result $a_{\mu}^{HAD}|_{uds}= 655\,(21)\,\times\,10^{-10}$ from \cite{LQCD3}, albeit obtained from a quite different method. 
After replacing the result Eq.(\ref{uds}) in Eq.(\ref{AMUHSF}) the leading order hadronic contribution determined entirely from theory is
%Eq.24
\begin{equation}
a_{\mu}^{HAD}|^{THY}_{QCD} = (729\, - \, 871)\,\times\,10^{-10}\;,\label{AMUF}
\end{equation}
which is significantly larger  than current estimates  using $e^+ e^-$ data, which lie in the range $a_{\mu}^{HAD} \simeq  \, (680 - 700) \,\times \, 10^{-10}$ \cite{review}. If the preliminary LQCD range \cite{Wittig}, Eq.(\ref{LQCDuds}), is confirmed, then the muon (g-2) value could be well understood within the Standard Model.
\section*{4. OPE-FESR analysis in the light-quark sector}
The basic idea of this approach in the light-quark region is to use Cauchy's theorem in the complex $s$-plane for the vector correlator, modulated by an analytic integration kernel designed to suppress the contribution from $e^+ e^-$ data. For convenience I begin by redefining the integration kernel in Eq.(\ref{AMUH1}) so that the anomaly is given by
%Eq.25
\begin{equation}
a_{\mu}^{\text{HAD},\text{LO}}=\int^{\infty}_{0}\tilde{K}(s)R(s)\,ds \;,\label{EQ:dispersion}
\end{equation}
where $\tilde{K}(s)$ is
%Eq.26
\begin{equation}
\tilde{K}(s) \equiv \frac{\alpha_{EM}^{2}}{3\,\pi^{2}} \frac{K(s)}{s}\,.
\end{equation}
Invoking Cauchy's theorem in the complex $s$-plane one finds
%Eq.27
\begin{equation}
\int^{s_0}_{0}p(s) R(s)\,ds - 6 \pi i \oint_{|s|=s_0}p(s) \Pi(s) \,ds = 0 \;, \label{eq:cauchy}
\end{equation}
where $p(s)$ is an arbitrary analytic function. Next, one invokes quark-hadron duality  to  replace $\Pi(s)$ by $\Pi_{\text{OPE}}(s)$ in the integral around the circle, where $\Pi_{\text{OPE}}(s)$ is the correlator in the framework of the OPE. The anomaly then becomes
%Eq.28
\begin{equation}
\tilde{a}_{\mu}^{\text{HAD},\text{LO}}(s_0)=\int^{s_0}_{0}\bigl[\tilde{K}(s)-p(s)\bigr]R(s)\,ds
+ 6\pi i\oint_{|s|=s_0}p(s)\;\Pi_{\text{OPE}}(s)\,ds  \;.\label{EQ:FESR}
\end{equation}
The next step is to choose an appropriate kernel $p(s)$ to suppress the contribution of the $e^+ e^-$ data in the relevant region $1\,\text{GeV}<\sqrt{s}<1.8\,\text{GeV}$, where it is badly known, and has the largest uncertainties. The optimal kernel was found to be \cite{amu2}
%Eq.29
\begin{equation}
p(s)=4.996\times 10^{-9} - 1.432\times 10^{-9} s \;,\label{EQ:fit}
\end{equation}
which minimizes the quantity
%Eq.30
\begin{equation}
\text{Max}\left|\frac{\tilde{K}(s)-p(s)}{\tilde{K}(s)}\right|, \ \  (1\,\text{GeV}<\sqrt{s}<1.8\,\text{GeV}). \label{EQ:MAX}
\end{equation}
This integration kernel $p(s)$ quenches all the data in the region mentioned above by at least a factor 2.5-3.0. The next step is to invoke the OPE in QCD entering the contour integral in Eq.(\ref{EQ:FESR})
%Eq.31
\begin{equation}
\Pi(q^2)\,=\, C_0\,\hat{I} \,+\,\sum_{N=0}\;C_{2N+2}(q^2,\mu^2)\;\langle\hat{O}_{2N+2}(\mu^2)   \rangle \;, \label{eq:OPE}
\end{equation}
where $\mu$ is a renormalization scale, and where the Wilson coefficients in this expansion, 
$ C_{2N+2}(q^2,\mu^2)$,  depend on the Lorentz indices and quantum numbers of $J(x)$ and  of the local gauge invariant operators $\hat{O}_N$ built from the quark and gluon fields. These operators are ordered by increasing dimensionality and the Wilson coefficients, calculable in PQCD, fall off by corresponding powers of $-q^2$. In other words, this OPE achieves a factorization of short distance effects encapsulated in the Wilson coefficients, and long distance dynamics present in the vacuum condensates. The unit operator $\hat{I}$ in Eq. \eqref{eq:OPE} has dimension $d=0$ and $C_0 \hat{I}$ stands for the purely PQCD contribution
%Eq.32
\begin{equation}
C_0 \hat{I} \equiv \Pi_{\text{PQCD}}(s)=\frac{Q_{T}^{2}}{16\pi^2}\left[\frac{20}{3} -  4 \log\left(-\frac{s}{\mu^2}\right)+\mathcal{O}(\alpha_{s})\right] \;.\label{OPEPQCD}
\end{equation}
Here $Q_{T}^{2}=\sum_{f=u,d,s} Q_{f}^{2}=2/3$. The perturbative correlator is known up to five-loop order  $\mathcal{O}(\alpha_s^4)$ \cite{chetyrkin1985}-\cite{baikov2008}. At dimension $d=2$ there are only quark-mass terms
%Eq.33
\begin{equation}
C_{2}\langle\mathcal{O}_{2}\rangle = \sum_{f=u,d,s}  \frac{Q_{f}^{2}}{4\pi^2} \; \frac{\bar{m}_{f}^{2}(\mu)}{s} \,\Bigl[6 + \mathcal{O}(\alpha_{s})\Bigr] \;,\label{EQ:masscorrection}
\end{equation}
where only the strange-quark makes a non-negligible contribution. At dimension $d=4$ the gluon  and the quark condensates contribute as
%Eq.34
\begin{equation}
C_{4}\langle\mathcal{O}_{4}\rangle= \frac{1}{s^2} \,\sum_{f=u,d,s} Q_{f}^{2} \Biggl\{\left[\frac{1}{12}-\frac{11}{216}\frac{\alpha_s (\mu)}{\pi}\right]\langle\frac{\alpha_s}{\pi}G^2\rangle 
+\left[2-\frac{2}{3}\frac{\alpha_s(\mu) }{\pi}+\mathcal{O}(\alpha_{s}^{2})\right]\overline{m}_f(\mu) \langle \bar{q_f} q_f\rangle(\mu)\Biggr\}\;. \label{C4O4}
\end{equation}
Finally, the QED contribution to the vector correlator needs to be included
%Eq.35
\begin{equation}
\Pi_{\text{QED}}(s)=\frac{3Q_{T}^{4}}{16\pi^2}\left[\frac{55}{12}-4 \,\zeta_3-\log\left(\frac{-s}{\mu ^2}\right)\right]\frac{\alpha_{\text{EM}}}{\pi} \;,
\end{equation} 
where $Q_{T}^{4}=\sum_{f=u,d,s} Q_{f}^{4}=2/9$.
This completes the information needed in the contour integral in Eq.(\ref{EQ:FESR}). The next step is to compute the line integral in Eq.(\ref{EQ:FESR}), which involves the data, for which we use the compilation in \cite{bodensteinOPE}. Regarding the numerical values of the parameters entering Eqs.(\ref{OPEPQCD})-(\ref{C4O4}), these are given in \cite{amu2}. The final result for the leading order hadronic contribution to the anomaly is
%Eq.36
\begin{equation}
\tilde{a}_{\mu}^{\text{HAD},\text{LO}}=(650.2\pm 4.0)\times 10^{-10} \;,\label{EQ:result4}
\end{equation}
where the error analysis may be found in \cite{amu2}. After adding the next-to-leading order contribution $a_{\mu}^{\text{HAD},\text{NLO}}=-10.1(6)\times 10^{-10}$, which has a different sign from 
$a_{\mu}^{\text{HAD},\text{LO}}$, the  prediction is
%Eq.37
\begin{equation}
\Delta a_\mu\equiv a_{\mu}^{\text{EXP}}-\tilde{a}_{\mu}^{\text{SM}}=19.2(8.0)\times 10^{-10}\;, \label{EQ:discr2}
\end{equation}
which is a lower $2.4 \,\sigma$ effect compared with the standard $3.3\, \sigma$ from $\Delta a_\mu\equiv a_{\mu}^{\text{EXP}}-a_{\mu}^{\text{SM}}=28.7(8.0)\times 10^{-10}$, using $e^+ e^-$ data for the leading order hadronic contribution. A detailed analysis of the validity of this approach may be found in \cite{amu2}.
%
%\section*{5. Conclusions}

%
\section*{Acknowledgements}
I wish to thank the Mainz Institute of Theoretical Physics, Johannes Gutenberg University of Mainz, Germany, for their generous support to this workshop. I also thank Hartmut Wittig and Hanno Horch for sharing their preliminary LQCD results on the derivative of the vector current correlator at the origin \cite{Wittig}. The fruitful collaboration with Sebastian Bodenstein in these projects is duly acknowledged.
% % % % % % % % % % % % % % % % % % % % % % % % % % % % % %
% % % % % % % % % % % % % % % % % % % % % % % % % % % % % %


\begin{thebibliography}{99}                      \begin{small}  
\bibitem{amu1} S. Bodenstein, C. A. Dominguez and K. Schilcher, Phys. Rev. D {\bf 85}, 014029 (2012).               %
\bibitem {review}J. P. Miller, E. de Rafael, and B. Lee Roberts, Rep. Prog.
Phys. \textbf{70}, 795 (2007); F. Jegerlehner, and A. Nyffeler, Phys. Rep.
\textbf{477}, 1 (2009),  and references therein.
\bibitem{LQCD1} B. Chakraborty {\it {et  al.}}, Phys. Rev. D \textbf {89}, 114501 (2014);
 J. Koponen {\it {et  al.}}, arXiv:1411.0569 (2014). 
\bibitem{LQCD2} B. Colquohoun {\it {et  al.}}, Phys. Rev. D \textbf{91}, 074514 (2015). 
\bibitem{LQCD3} F. Burger {\it {et  al.}}, J. High Ener. Phys. \textbf{1402}, 099 (2014)
\bibitem{Wittig} H. Wittig, private communication.
\bibitem{amu2}S. Bodenstein, C. A. Dominguez,  K. Schilcher and H. Spiesberger Phys. Rev. D {\bf 88}, 014005 (2013). 
\bibitem{DV} D. Boito, M. Golterman, K. Maltman, J. Osborne, and S. Peris, Phys. Rev. D {\bf 91} (2015) 034003.
\bibitem{Cata} O. Cat\`a, M. Golterman, and S. Peris, Phys. Rev. D {\bf 79}, 053002 (2009).
\bibitem {deRK} S. J. Brodsky and E. de Rafael, Phys. Rev. \textbf{168}, 1620 (1968).
\bibitem{5L} P. A. Baikov, K. G. Chetyrkin, and J. H. K\"{u}hn, Phys. Rev.
Lett. \textbf{96}, 012003 (2006).
\bibitem{BES} J. Z. Bai {\it et al.}, BES Coll., Phys. Rev. Lett. {\bf 88}, 101802 (2002);
M. Ablikim {\it et al.}, BES Coll., Phys. Lett. B {\bf 677}, 239 (2009).
\bibitem {boughezal2006a}R. Boughezal, M. Czakon, and T. Schutzmeier, Phys.
Rev. D \textbf{74}, 074006 (2006).
\bibitem {maier2008a}A. Maier, P. Maierh\"{o}fer, and P. Marquard, Nucl. Phys.
B \textbf{797}, 218 (2008); Phys. Lett. B \textbf{669}, 88 (2008).
\bibitem {chetyrkin2006}K. G. Chetyrkin, J. H. K\"{u}hn, and C. Sturm, Eur.
Phys. J. C \textbf{48}, 107 (2006).
\bibitem {maier2010}A. Maier \textit{et al.}, Nucl. Phys. B \textbf{824}, 1 (2010).
\bibitem{Chetyrkin1997} K. G. Chetyrkin \textit{et al.}, Nucl. Phys. B \textbf{503}, 339 (1997).
\bibitem{MAIER2} A. Maier, and P. Marquard, arXiv:1110.558.
\bibitem {Baikov2009}P. A. Baikov, K. G. Chetyrkin, and J. H. K\"{u}hn, Nucl.
Phys. B (Proc. Suppl.) \textbf{189}, 49 (2009).
\bibitem {Chetyrkin2000b}K. G. Chetyrkin, R. Harlander, and J. H. K\"{u}hn,
Nucl. Phys. B \textbf{586}, 56 (2000).
\bibitem {Baikov2008}P. A. Baikov, K. G. Chetyrkin, and J. H. K\"{u}hn, Phys.
Rev. Lett. \textbf{101}, 012002 (2008).
\bibitem {Baikov2004}P. A. Baikov, K. G. Chetyrkin, and J. H. K\"{u}hn, Nucl.
Phys. B (Proc. Suppl.) \textbf{135}, 243 (2004).
%\bibitem{PIFF}C. A. Dominguez, Phys. Lett. B {\bf{512}}, 331 (2001).
%\bibitem{NFF} C. A. Dominguez and T. Thapedi, J.  High Ener. Phys.  
%{\bf {0410}}, 003 (2004).
%\bibitem{DeltaFF} C. A. Dominguez and R. R\"{o}ntsch, J. High Ener. Phys.  
%{\bf {0710}}, 085 (2007).
%\bibitem{AMEN} C. J. Bebek {\it et al.}, Phys. Rev. D {\bf{17}}, 1693 %(1978); S. R. Amendolia {\it et al.}, Nucl. Phys. B {\bf{277}}, 168 %(1986) ; J. Volmer {\it et al.}, Phys. Rev. Lett. {\bf{86}},  1713 %(2001).
\bibitem{chetyrkin1985}  K. G. Chetyrkin, V. P. Spiridonov and S. G. Gorishnii, Phys. Lett. B {\bf 160}, 149 (1985).
\bibitem{gorishnii1991}  S. Gorishnii, A. Kataev and S. Larin, Phys. Lett. B {\bf 259}, 144 (1991).
\bibitem{surguladze1991}  L.R. Surguladze and M.A. Samuel, Phys. Rev. Lett. {\bf 66}, 560 (1991)Erratum, \emph{ibid}. {\bf 66}, 2416 (1991).
\bibitem{chetyrkin1979}  K. Chetyrkin, A. Kataev and F. Tkachov, Phys. Lett. B {\bf 85}, 277 (1979).
\bibitem{dine1979}  M. Dine and J. Sapirstein, Phys. Rev. Lett. {\bf 43}, 668 (1979).
\bibitem{celmaster1980}  W. Celmaster and R.J. Gonsalves, Phys. Rev. Lett. {\bf 44}, 560 (1980).
\bibitem{baikov2008}  P. Baikov, K. Chetyrkin and J.H. Kuhn, Phys. Rev. Lett. {\bf 101}, 012002 (2008).
\bibitem{bodensteinOPE}  S. Bodenstein, C.A. Dominguez, S.I. Eidelman, H. Spiesberger and K. Schilcher, JHEP {\bf 1}, 39 (2012).
\end{small} 
\end{thebibliography}
\end{document}